# The dual formalisms of nonextensive thermodynamics for open systems with maximum entropy principle

Yahui Zheng[1], Haining Yu[2], Jiulin Du[3]

1. *Department of Electronics and Communication Engineering, Henan Institute of Technology, Xinxiang City 453003, China.*

2. *Shanxi Institute of Energy, Jinzhong City 030600, China.*

3. *Department of Physics, School of Science, Tianjin University, Tianjin 300072, China.*

## Abstract

We study the nonextensive thermodynamics for open systems. On the basis of the maximum entropy principle, the dual power-law $q$-distribution functions are re-deduced by using the dual particle number definitions and assuming that the chemical potential is constant in the two sets of parallel formalisms, where the fundamental thermodynamic equations with dual interpretations of thermodynamic quantities are derived for the open systems. By introducing parallel structures of Legendre transformations, other thermodynamic equations with dual interpretations of quantities are also deduced in the open systems, and then several dual thermodynamic relations are inferred. One can easily find that there are correlations between the dual relations, from which an equivalent rule is found that the Tsallis factor is invariable in calculations of partial derivative with constant volume or constant entropy. Using this rule, more correlations can be found. And the statistical expressions of the Lagrange internal energy and pressure are easily obtained.

**Keywords:** nonextensive thermodynamic equation; nonextensive thermodynamic relation; Legendre transformation

**PACS: 05.70.Ce; 05.90.+m; 51.30.+i.**

## 1. Introduction

In recent years, nonextensive statistical mechanics (NSM) has been developed, based on the $q$-entropy firstly proposed by Tsallis in 1988 [1]. Different from the classical statistics, this new paradigm of statistical theory suggested a type of power-law distribution functions, which have been applied into many interesting research fields, such as self-gravitating astrophysical systems [2-6], astrophysical and space plasmas [7-10], chemical reaction systems [11,12], biological systems [13-15] and so on. Due to the success in dealing with the non-Maxwellian distribution functions of complex systems, NSM is being popularly accepted and applied by many authors in very wide science and technology fields.

On the other hand, because the fundamental laws of thermodynamics were originated from the phenomenological observations and experiments, people always expect that they can be appropriate to any system, no matter what their statistical basis is. This is reasonable in logic, yet very difficult in the real theoretical development of the thermodynamics in NSM. For example, in order to define an appropriate pressure, many authors [16-19] abandoned the standard Legendre transformation and suggested a modified free energy where the Tsallis factor [20] occurred in an obvious way. This results in a direct obstacle to the further introductions of other Legendre transformations.

In order to develop the nonextensive thermodynamic formalism with the same mathematical structure (i.e., the same thermodynamic relations) as the traditional one, we proposed the dual interpretations of the physical quantities and fundamental thermodynamic equations based on the temperature duality assumption [20]. And then, the nonextensive thermodynamic formalism consisting of two parallel Legendre transformation structures was presented. One is called the physical set (P-set), the other is called the Lagrange set (L-set). In each set of thermodynamic formalism, the Legendre transformation is obtained directly from the classical thermodynamics. The only difference is that there are two sets of transformation structures in our treatment. The Tsallis factor does not appear in any set of the formalism, but these two sets of formalisms are linked through the Tsallis factor. It can be seen that thermodynamic relations in the proposed thermodynamics are almost the same as those in traditional thermodynamics and therefore can be applied to any realistic systems. The key point is that the quantities in complex systems should be interpreted in the dual ways.

In this work, we would further study the nonextensive thermodynamic equations in open systems on basis of the maximum entropy principle, where particle number is also a state variable. The remaining of this paper is constructed as follows. In section 2, the dual power-law distribution functions are re-deduced by recourse to the maximum entropy principle. In section 3, the first and the zeroth laws of thermodynamics are revisited for the open systems. In section 4, the links between the fundamental thermodynamic equations in nonextensive realm and the balance conditions are studied. In section 5, the Legendre transformations are introduced and the corresponding nonextensive thermodynamic relations are deduced for the open systems. In section 6, the correlations of the nonextensive thermodynamic relations within different set of formalisms for the open systems are found. In section 7, we give the conclusions and discussions.

## 2. The dual power-law $q$-distribution functions with maximum entropy principle

Firstly, let us start from the Tsallis q-entropy, expressed [1] by

$$S_q = k \frac{\sum p_i^q - 1}{1 - q} \equiv k \sum p_i \ln_q(\frac{1}{p_i}),$$  (1)

where $k$ is the Boltzmann constant, $p_i$ is the probability of the $i$th microstate of the system and $q$ is the nonextensive parameter. For convenience, we define $c_q = \sum p_i^q$ as Tsallis factor. In the nonextensive thermodynamics, two different internal energies were proposed [20]. One is the physical internal energy (P-internal energy), defined as [21]

$$U_q^{(2)} \equiv \sum p_i^q \varepsilon_i,$$  (2)

the other is the Lagrange internal energy (L-internal energy), defined as [22]

$$U_q^{(3)} \equiv \frac{\sum p_i^q \varepsilon_i'}{c_q}.$$  (3)

Here, we should notice that the microstate energy levels, $\varepsilon_i$ and $\varepsilon_i'$, in above two definitions are different from each other. The main difference between them is that they obey different composite rules, which will be discussed later on. It is apparent that these two definitions (2) and (3) have been employed in literature. However, in the present work we would endow them with new physical senses.

On the other hand, because the particle number is a variable in an open system, here we propose the dual interpretations of averaged particle number. That is the physical particle number (P-number),

$$\bar{N}^{(2)} \equiv \sum p_i^q N_i \,, \tag{4}$$

and the Lagrange particle number (L-number),

$$\bar{N}^{(3)} \equiv \frac{\sum p_i^q N_i'}{c_q} \,. \tag{5}$$

Again we notice that the microstate particle numbers, $N_i$ and $N_i'$, in the two definitions are also different from each other. Similarly, we will see later on that they obey different composite rules.

In order to apply the maximum entropy principle, we define the functional in P-set of formalism for the open system as

$$F^{(2)}(\{p_i\}) \equiv \frac{S_q}{k} - \alpha(\sum p_i - 1) - \beta'(\sum p_i^q \varepsilon_i - U_q^{(2)}) - \gamma'(\sum p_i^q N_i - \bar{N}^{(2)}) \,, \tag{6}$$

where the quantity $\alpha$ is a multiplier related to the normalization of probability, the quantity $\beta'$ is the generalized Lagrange multiplier and the $\gamma'$ is also a generalized multiplier related to chemical potential. According to the maximum entropy principle, the extreme value of the above function (6) should be zero, which leads to the equation [23],

$$\frac{q}{1-q} \sum p_i^{q-1} \delta p_i - \alpha \sum \delta p_i - \beta' q \sum p_i^{q-1} \varepsilon_i \delta p_i - \gamma' q \sum p_i^{q-1} N_i \delta p_i = 0 \,. \tag{7}$$

According to the variation principle, we have that

$$\frac{q}{1-q} p_i^{q-1} - \beta' q p_i^{q-1} \varepsilon_i - \gamma' q p_i^{q-1} N_i = \alpha \,. \tag{8}$$

Then the power-law distribution function is obtained,

$$p_i^{(2)} = \frac{1}{Z_q^{(2)}} [1 - (1-q)(\beta' \varepsilon_i + \gamma' N_i)]^{1/(1-q)} \,,$$

$$Z_q^{(2)} \equiv \sum [1 - (1-q)(\beta' \varepsilon_i + \gamma' N_i)]^{1/(1-q)} \,. \tag{9}$$

It is easy to find that [23]

$$U_q^{(2)} = -\frac{\partial}{\partial \beta'} \ln_q Z_q^{(2)} \,, \tag{10}$$

$$\bar{N}^{(2)} = -\frac{\partial}{\partial \gamma'} \ln_q Z_q^{(2)} \,. \tag{11}$$

And furthermore we can find that [24]

$$c_q = [Z_q^{(2)}]^{1-q} + (1-q)(\beta' U_q^{(2)} + \gamma' \bar{N}^{(2)}) \,. \tag{12}$$

Similarly, in order to apply the maximum entropy principle to the open system in the L-set of formalism, the function can be written by,

$$F^{(3)}(\{p_i\}) = \frac{S_q}{k} - \alpha(\sum p_i - 1) - \beta(\frac{\sum p_i^q \varepsilon_i'}{c_q} - U_q^{(3)}) - \gamma(\frac{\sum p_i^q N_i'}{c_q} - \bar{N}^{(3)}) \,, \tag{13}$$

where the quantity $\beta$ is the Lagrange multiplier, and the $\gamma$ is a multiplier related to the chemical potential. According to the variation principle, the extreme value of the above function (13) also leads to the power-law distribution,

$$p_i^{(3)} = \frac{1}{\overline{Z}_q^{(3)}}[1 - \frac{(1-q)}{c_q}[\beta(\varepsilon_i' - U_q^{(3)}) + \gamma(N_i - \overline{N}^{(3)})]]^{1/(1-q)} ,$$

$$\overline{Z}_q^{(3)} \equiv \sum [1 - \frac{(1-q)}{c_q}[\beta(\varepsilon_i' - U_q^{(3)}) + \gamma(N_i - \overline{N}^{(3)})]]^{1/(1-q)} . \qquad (14)$$

It is difficult to deduce the statistical expressions of L-internal energy and L-number directly, yet we can prove that (see the Appendix A) [22]

$$c_q = [\overline{Z}_q^{(3)}]^{1-q} . \qquad (15)$$

In next section, by recourse to the power-law distribution functions presented in this section we would discuss the first and the zeroth laws of thermodynamics for the open systems.

## 3. The first and zeroth laws of nonextensive thermodynamics for open systems

In order to rebuild the first law of thermodynamics in the domain of P-set of formalism for the open systems in nonextensive thermodynamics, we should both consider the variations of internal energy and the averaged particle number, namely,

$$dU_q^{(2)} = \sum p_i^q d\varepsilon_i + q \sum p_i^{q-1} \varepsilon_i dp_i , \text{ and} \qquad (16)$$

$$d\overline{N}^{(2)} = \sum p_i^q dN_i + q \sum p_i^{q-1} N_i dp_i , \qquad (17)$$

from which we can deduce that

$$\frac{\beta' dU_q^{(2)} + \gamma' d\overline{N}^{(2)}}{\beta'} = \sum p_i^q \frac{(\beta' d\varepsilon_i + \gamma' dN_i)}{\beta'} + q \sum p_i^{q-1} \frac{(\beta' \varepsilon_i + \gamma' N_i)}{\beta'} dp_i . \qquad (18)$$

For an open system, the action exerted by external force can lead to the shift in energy level and the change in particle number on each microstate at the same time. Therefore, it is reasonable to judge that the first term on the right hand side of the above equality (18) is the work done by the reservoir. Similarly, in open system the heat conduction process always changes the distribution function of system with fixed microstate energy level and particle number. So the second term on the right hand side of (18) is the heat absorbed by the system from the reservoir.

Then, we have the work

$$dW^{(2)} = \sum p_i^q (d\varepsilon_i + \frac{\gamma'}{\beta'} dN_i) = -P_q dV , \qquad (19)$$

where the $P_q$ is the P-pressure. We also has the heat

$$dQ^{(2)} = q \sum p_i^{q-1} \frac{(\beta' \varepsilon_i + \gamma' N_i)}{\beta'} dp_i$$

$$= \frac{q[Z_q^{(2)}]^{1-q}}{(1-q)\beta'} \sum [1 - (1-q)(\beta' \varepsilon_i + \gamma' N_i)]^{-1} dp_i . \qquad (20)$$

On the other hand, the variation of the $q$-entropy gives out that

$$dS_q = k \frac{q[Z_q^{(2)}]^{1-q}}{1-q} \sum [1 - (1-q)(\beta' \varepsilon_i + \gamma' N_i)]^{-1} dp_i , \qquad (21)$$

which leads to

$$dQ^{(2)} = \frac{dS_q}{k\beta'} . \qquad (22)$$

Now we introduce the generalized Lagrange relations,

$$\beta' = \frac{1}{kT_q}, \quad \gamma' = -\mu\beta' = -\frac{\mu}{kT_q}, \tag{23}$$

where the $T_q$ is the P-temperature and the $\mu$ is the chemical potential. Here we generally suppose that the chemical potential is identical in these two sets of formalisms.

According to the relations in (23), we get

$$dU_q^{(2)} = T_q dS_q - P_q dV + \mu d\bar{N}^{(2)}, \tag{24}$$

which is the fundamental nonextensive thermodynamic equation for the open systems in P-set of formalism. In order to calculate the one in L-set of formalism, we consider the variations of the corresponding L-internal energy and L-number, that is to say, [18]

$$dU_q^{(3)} = \frac{\sum p_i^q d\varepsilon_i'}{c_q} + q\frac{\sum p_i^{q-1}(\varepsilon_i' - U_q^{(3)})dp_i}{c_q} \quad \text{and} \tag{25}$$

$$d\bar{N}^{(3)} = \frac{\sum p_i^q dN_i'}{c_q} + q\frac{\sum p_i^{q-1}(N_i' - \bar{N}^{(3)})dp_i}{c_q}, \tag{26}$$

which leads to

$$\frac{\beta dU_q^{(3)} + \gamma d\bar{N}^{(3)}}{\beta} = \frac{\sum p_i^q(\beta d\varepsilon_i' + \gamma dN_i')}{\beta c_q} + q\frac{\sum p_i^{q-1}[\beta(\varepsilon_i' - U_q^{(3)}) + \gamma(N_i' - \bar{N}^{(3)})]dp_i}{\beta c_q}. \tag{27}$$

Likewise, for open system, the external force changes microstate energy level and particle number, but the heat conduction changes the system distribution function with unchanged energy level and particle number. So we confirm that the first term on the right-hand side of the above equation (27) is the work done by the reservoir, and the second term is the heat absorbed by the system. So there are

$$dW^{(3)} = \frac{\sum p_i^q(d\varepsilon_i' + \gamma dN_i'/\beta)}{c_q} = -PdV, \tag{28}$$

$$dQ^{(3)} = q\frac{\sum p_i^{q-1}[\beta(\varepsilon_i' - U_q^{(3)}) + \gamma(N_i' - \bar{N}^{(3)})]dp_i}{\beta c_q}$$

$$= \frac{c_q q}{(1-q)\beta}\sum[1 - \frac{(1-q)}{c_q}[\beta(\varepsilon_i' - U_q^{(3)}) + \gamma(N_i - \bar{N}^{(3)})]]^{-1}dp_i, \tag{29}$$

where the $P$ is the L-pressure.

Moreover, the variation of $q$-entropy in the L-set of formalism can be written as [18]

$$dS_q = k\frac{c_q q}{1-q}\sum[1 - \frac{(1-q)}{c_q}[\beta(\varepsilon_i' - U_q^{(3)}) + \gamma(N_i - \bar{N}^{(3)})]]^{-1}dp_i. \tag{30}$$

Then we can find that

$$dQ^{(3)} = \frac{dS_q}{k\beta}. \tag{31}$$

Now let us introduce the following Lagrange relations,

$$\beta = \frac{1}{kT}, \quad \gamma = -\mu\beta = -\frac{\mu}{kT}, \tag{32}$$

where the $T$ is the L-temperature. According to the relations in (32), we obtain the fundamental thermodynamic equation for the open systems in the L-set of formalism, namely,

$$dU_q^{(3)} = TdS_q - PdV + \mu d\bar{N}^{(3)}. \tag{33}$$

In order to make clear the physical meanings of the temperatures in (24) and (33), we need to revisit the zeroth law of thermodynamics in each set of formalisms. For this aim, we should know the addition rules of the microstate energy levels and the particle numbers. Noticing that they are different in each set of the formalisms, we generally suggest the following composite rules

$$\varepsilon_{(i,j)(1,2)} - \mu N_{(i,j)(1,2)} = \varepsilon_{i1} + \varepsilon_{j2} - \mu N_{i1} - \mu N_{j2} - (1-q)\beta'(\varepsilon_{i1} - \mu N_{i1})(\varepsilon_{j2} - \mu N_{j2}), \qquad (34)$$

$$\varepsilon'_{(i,j)(1,2)} - \mu N'_{(i,j)(1,2)} = \varepsilon'_{i1} + \varepsilon'_{j2} - \mu N'_{i1} - \mu N'_{i2}$$

$$-(1-q)\frac{\beta}{c_q}[(\varepsilon'_{i1} - U^{(3)}_{q1}) - \mu(N'_{i1} - \bar{N}^{(3)}_{1})][(\varepsilon'_{i2} - U^{(3)}_{q2}) - \mu(N'_{i2} - \bar{N}^{(3)}_{2})], \qquad (35)$$

where "1" and "2" represent different subsystems; "i" and "j" represent different microscopic states. It can be proved that the above two addition rules guarantee the validity of the probability independence assumption, $p_{ij} = p_i p_j$.

According to the probability independence assumption, there are the following addition rules of macroscopic physical quantities in nonextensive thermodynamics, that is,

$$S_{q(1,2)} = S_{q1} + S_{q2} + \frac{1-q}{k}S_{q1}S_{q2}, \qquad (36)$$

$$U^{(2)}_{q(1,2)} - \mu \bar{N}^{(2)}_{(1,2)} = c_{q2}(U^{(2)}_{q1} - \mu \bar{N}^{(2)}_{1}) + c_{q1}(U^{(2)}_{q2} - \mu \bar{N}^{(2)}_{2})$$

$$-(1-q)\beta'(U^{(2)}_{q1} - \mu \bar{N}^{(2)}_{1})(U^{(2)}_{q2} - \mu \bar{N}^{(2)}_{2}), \qquad (37)$$

$$U^{(3)}_{q(1,2)} - \mu \bar{N}^{(3)}_{(1,2)} = U^{(3)}_{q1} - \mu \bar{N}^{(3)}_{1} + U^{(3)}_{q2} - \mu \bar{N}^{(3)}_{2}. \qquad (38)$$

It can be seen that the $q$-entropy, the P-internal energy, and the P-number are all nonadditive. On the contrary, now that the chemical potential is indefinite, the equation (38) suggests the additive L-internal energy and L-number, i.e.,

$$U^{(3)}_{q(1,2)} = U^{(3)}_{q1} + U^{(3)}_{q2}, \quad \bar{N}^{(3)}_{(1,2)} = \bar{N}^{(3)}_{1} + \bar{N}^{(3)}_{2}. \qquad (39)$$

We regard the subsystem 1 as the researched object, and the subsystem 2 as its reservoir. Between these two subsystems there exists the exchange of internal energy and particle number. Now we consider the variations of (36), (37) and (39), respectively,

$$\delta S_{q(1,2)} = c_{q2}\delta S_{q1} + c_{q1}\delta S_{q2}, \qquad (40)$$

$$\delta U^{(2)}_{q(1,2)} - \mu \delta \bar{N}^{(2)}_{(1,2)} = c_{q2}(\delta U^{(2)}_{q1} - \mu \delta \bar{N}^{(2)}_{1}) + c_{q1}(\delta U^{(2)}_{q2} - \mu \delta \bar{N}^{(2)}_{2}), \qquad (41)$$

$$\delta U^{(3)}_{q(1,2)} = \delta U^{(3)}_{q1} + \delta U^{(3)}_{q2}, \qquad (42)$$

$$\delta \bar{N}^{(3)}_{(1,2)} = \delta \bar{N}^{(3)}_{1} + \delta \bar{N}^{(3)}_{2}. \qquad (43)$$

In the calculation of (41), the relation (12) is taken into account. Similarly, now that the chemical potential is indefinite, the equation (41) actually suggests two independent nonadditive relations, that is,

$$\delta U^{(2)}_{q(1,2)} = c_{q2}\delta U^{(2)}_{q1} + c_{q1}\delta U^{(2)}_{q2}, \qquad (44)$$

$$\delta \bar{N}^{(2)}_{(1,2)} = c_{q2}\delta \bar{N}^{(2)}_{1} + c_{q1}\delta \bar{N}^{(2)}_{2}. \qquad (45)$$

When the whole system arrive the $q$-equilibrium state, taking into account

$$\delta S_{q(1,2)} = \delta U_{q(1,2)}^{(3)} = 0 ,\tag{46}$$

we gets

$$c_{q1}T_1 = c_{q2}T_2 .\tag{47}$$

On the other hand, letting

$$\delta S_{q(1,2)} = \delta U_{q(1,2)}^{(2)} = 0 ,\tag{48}$$

we obtains

$$T_{q1} = T_{q2} .\tag{49}$$

Now that the equations (47) and (49) denote the same $q$-equilibrium state, generally, we have

$$T_q = c_q T .\tag{50}$$

This relation is called the assumption of temperature duality [20], which at the same time gives the physical reality to the P-temperature and L-temperature. Here, we should emphasize that although these two temperatures are both real in physics, the P-temperature is related to the global property of the system, such as its balance condition and evolution process, while the L-temperature is related to the local nature of system, such as the local thermal equilibrium. Therefore, in experiments, the P-temperature is un-measureable, while the L-temperature is measurable. In next section, we will study the links of the basic nonextensive thermodynamic equations and analyze the balance conditions for the $q$-equilibrium.

## 4. The links of the nonextensive thermodynamic equations to the balance conditions

Consider these two works given in (19) and (28) respectively, if neglecting the difference between the microstate energy levels and the particle numbers, we have

$$dW^{(2)} = c_q dW^{(3)},\tag{51}$$

leading to the pressure link

$$P_q = c_q P .\tag{52}$$

In view of (50) and (52), from (18) and (27) we get

$$dU_q^{(2)} - \mu d\bar{N}^{(2)} = c_q[dU_q^{(3)} - \mu d\bar{N}^{(3)}] .\tag{53}$$

Comparing (43) and (45), we find

$$d\bar{N}^{(2)} = c_q d\bar{N}^{(3)} ,\tag{54}$$

which directly results in the link of these two fundamental nonextensive thermodynamic equations (24) and (33), that is,

$$dU_q^{(2)} = c_q dU_q^{(3)} .\tag{55}$$

Moreover, if ignoring the difference between the microstate energy levels in (2) and (3), and ignoring the difference between the microstate particle numbers in (4) and (5), we also have

$$U_q^{(2)} = c_q U_q^{(3)}, \quad \bar{N}^{(2)} = c_q \bar{N}^{(3)} .\tag{56}$$

In the deductions of (37) and (38), we have assumed the generalized Lagrange multiplier $\beta'$ and the chemical potential to be constant. This means that the invariable P-temperature and the invariable chemical potential are the balance conditions for the $q$-equilibrium. Apart from these two quantities, the invariable P-pressure is also the balance condition. In order to prove this, we assume [25] that the volume is also nonadditive and satisfies

$$\delta V_{(1,2)} = c_{q2}\delta V_1 + c_{q1}\delta V_2 \, . \tag{57}$$

Then in the P-set of formalism, in view of

$$dS_q = \frac{dU_q^{(2)} + P_q dV - \mu d\bar{N}^{(2)}}{T_q} \, , \tag{58}$$

and taking into account for the $q$-equilibrium that

$$\delta S_{q(1,2)} = \delta U_{q(1,2)}^{(2)} = \delta \bar{N}_{(1,2)}^{(2)} = \delta V_{(1,2)} = 0 \, , \tag{59}$$

we can obtain

$$c_{q2}(\frac{1}{T_{q1}} - \frac{1}{T_{q2}})dU_{q1}^{(2)} + c_{q2}(\frac{P_{q1}}{T_{q1}} - \frac{P_{q2}}{T_{q2}})dV_1 - c_{q2}(\frac{\mu_1}{T_{q1}} - \frac{\mu_2}{T_{q2}})d\bar{N}_1^{(2)} = 0 \, . \tag{60}$$

Furthermore, in view of

$$dS_q = \frac{dU_q^{(3)} + PdV - \mu d\bar{N}^{(3)}}{T} \, , \tag{61}$$

and letting

$$\delta S_{q(1,2)} = \delta U_{q(1,2)}^{(3)} = \delta \bar{N}_{(1,2)}^{(3)} = \delta V_{(1,2)} = 0 \, , \tag{62}$$

we can obtain

$$(\frac{c_{q2}}{T_1} - \frac{c_{q1}}{T_2})dU_{q1}^{(3)} + c_{q2}(\frac{P_1}{T_1} - \frac{P_2}{T_2})dV_1 - (\frac{c_{q2}\mu_1}{T_1} - \frac{c_{q1}\mu_2}{T_2})d\bar{N}_1^{(3)} = 0 \, . \tag{63}$$

It is obvious that, in light of the links in (50), (52), (54) and (55), the equations (60) and (63) produce the same balance conditions, namely,

$$T_{q1} = T_{q2}, \quad P_{q1} = P_{q2}, \quad \mu_1 = \mu_2 \, . \tag{64}$$

It should be noticed that in the P-set of thermodynamic formalism, the balance conditions are deduced more directly. Although the inference in the conditions (64) is dependent on the assumption (57) of the volume nonadditivity, the latter is compatible with other thermodynamic relations. Actually, if take the conditions (64) as the logic start point, the assumption (57) can be inferred easily. In next section, we would discuss the Legendre transformations for the open systems.

## 5. The Legendre transformations and the nonextensive thermodynamic relations for open systems

In our treatment, all the Legendre transformations can be directly derived from the classical thermodynamics in each set of nonextensive thermodynamic formalisms. Therefore we can directly write down the free energies, enthalpies, and Gibbs functions, as follows,

$$F_q^{(2)} = U_q^{(2)} - T_q S_q \, , \tag{65}$$

$$F_q^{(3)} = U_q^{(3)} - T S_q \, , \tag{66}$$

$$H_q^{(2)} = U_q^{(2)} + P_q V \, , \tag{67}$$

$$H_q^{(3)} = U_q^{(3)} + P V \, , \tag{68}$$

$$G_q^{(2)} = U_q^{(2)} - T_q S_q + P_q V \, , \tag{69}$$

$$G_q^{(3)} = U_q^{(3)} - T S_q + P V \, . \tag{70}$$

It is easy to find the links of these two sets of fundamental thermodynamic functions, according to the fundamental links in equations (50), (52), and (56), namely,

$$F_q^{(2)} = c_q F_q^{(3)}, \tag{71}$$

$$H_q^{(2)} = c_q H_q^{(3)}, \tag{72}$$

$$G_q^{(2)} = c_q G_q^{(3)}. \tag{73}$$

Besides, according to (12), the statistical expression of the P-free energy can be written as

$$F_q^{(2)} = -kT_q \ln_q Z_q^{(2)} + \mu \bar{N}^{(2)}, \tag{74}$$

which is different from the P-free energy expression in a closed system [20] with constant particle number. Actually, just as the traditional thermodynamics, we can further introduce the so-called grand thermodynamic potential $J$ in each set of formalisms, that is,

$$J_q^{(2)} = F_q^{(2)} - \mu \bar{N}^{(2)}, \tag{75}$$

$$J_q^{(3)} = F_q^{(3)} - \mu \bar{N}^{(3)}. \tag{76}$$

And then there is

$$J_q^{(2)} = c_q J_q^{(3)}. \tag{77}$$

The statistical expression of the P-grand potential is obvious, according to (74),

$$J_q^{(2)} = -kT_q \ln_q Z_q^{(2)}, \tag{78}$$

which is similar to the expression of the grand in traditional thermodynamics.

Based on the Legendre transformations mentioned above, we symmetrically write down the differential expressions of the thermodynamic functions in each set of the nonextensive formalisms, namely,

$$dU_q^{(2)} = T_q dS_q - P_q dV + \mu d\bar{N}^{(2)}, \tag{79}$$

$$dU_q^{(3)} = T dS_q - P dV + \mu d\bar{N}^{(3)}, \tag{80}$$

$$dF_q^{(2)} = -P_q dV - S_q dT_q + \mu d\bar{N}^{(2)}, \tag{81}$$

$$dF_q^{(3)} = -P dV - S_q dT + \mu d\bar{N}^{(3)}, \tag{82}$$

$$dH_q^{(2)} = T_q dS_q + V dP_q + \mu d\bar{N}^{(2)}, \tag{83}$$

$$dH_q^{(3)} = T dS_q + V dP + \mu d\bar{N}^{(3)}, \tag{84}$$

$$dG_q^{(2)} = -S_q dT_q + V dP_q + \mu d\bar{N}^{(2)}, \tag{85}$$

$$dG_q^{(3)} = -S_q dT + V dP + \mu d\bar{N}^{(3)}, \tag{86}$$

$$dJ_q^{(2)} = -P_q dV - S_q dT_q - \bar{N}^{(2)} d\mu, \tag{87}$$

$$dJ_q^{(3)} = -P dV - S_q dT - \bar{N}^{(3)} d\mu. \tag{88}$$

By recourse to the partial derivative theory, it is easy to prove the Maxwellian relations, now that the Legendre transformations are directly derived from the traditional thermodynamics. Here we write down the Maxwellian relations in the L-set only, and those in P-set can be obtained directly by the simple symbol substitution. These Maxwellian relations in L-set are written as

$$\left( \frac{\partial T}{\partial V} \right)_{S_q, \bar{N}^{(3)}} = -\left( \frac{\partial P}{\partial S_q} \right)_{V, \bar{N}^{(3)}}, \tag{89}$$

$$\left( \frac{\partial T}{\partial \bar{N}^{(3)}} \right)_{S_q, V} = \left( \frac{\partial \mu}{\partial S_q} \right)_{V, \bar{N}^{(3)}}, \left( \frac{\partial P}{\partial \bar{N}^{(3)}} \right)_{V, S_q} = -\left( \frac{\partial \mu}{\partial V} \right)_{S_q, \bar{N}^{(3)}}, \tag{90}$$

$$\left(\frac{\partial S_q}{\partial V}\right)_{T,\bar{N}^{(3)}} = \left(\frac{\partial P}{\partial T}\right)_{V,\bar{N}^{(3)}}, \tag{91}$$

$$\left(\frac{\partial S_q}{\partial \bar{N}^{(3)}}\right)_{T,V} = -\left(\frac{\partial \mu}{\partial T}\right)_{V,\bar{N}^{(3)}}, \left(\frac{\partial P}{\partial \bar{N}^{(3)}}\right)_{T,V} = -\left(\frac{\partial \mu}{\partial V}\right)_{T,\bar{N}^{(3)}}, \tag{92}$$

$$\left(\frac{\partial T}{\partial P}\right)_{S_q,\bar{N}^{(3)}} = \left(\frac{\partial V}{\partial S_q}\right)_{P,\bar{N}^{(3)}}, \tag{93}$$

$$\left(\frac{\partial T}{\partial \bar{N}^{(3)}}\right)_{S_q,P} = \left(\frac{\partial \mu}{\partial S_q}\right)_{P,\bar{N}^{(3)}}, \left(\frac{\partial V}{\partial \bar{N}^{(3)}}\right)_{S_q,P} = \left(\frac{\partial \mu}{\partial P}\right)_{S_q,\bar{N}^{(3)}}, \tag{94}$$

$$\left(\frac{\partial S_q}{\partial P}\right)_{T,\bar{N}^{(3)}} = -\left(\frac{\partial V}{\partial T}\right)_{P,\bar{N}^{(3)}}, \tag{95}$$

$$\left(\frac{\partial S_q}{\partial \bar{N}^{(3)}}\right)_{T,P} = -\left(\frac{\partial \mu}{\partial T}\right)_{P,\bar{N}^{(3)}}, \left(\frac{\partial V}{\partial \bar{N}^{(3)}}\right)_{T,P} = \left(\frac{\partial \mu}{\partial P}\right)_{T,\bar{N}^{(3)}}, \tag{96}$$

$$\left(\frac{\partial P}{\partial T}\right)_{V,\mu} = \left(\frac{\partial S_q}{\partial V}\right)_{T,\mu}, \tag{97}$$

$$\left(\frac{\partial P}{\partial \mu}\right)_{V,T} = \left(\frac{\partial \bar{N}^{(3)}}{\partial V}\right)_{T,\mu}, \left(\frac{\partial S_q}{\partial \mu}\right)_{V,T} = \left(\frac{\partial \bar{N}^{(3)}}{\partial T}\right)_{V,\mu}. \tag{98}$$

Furthermore, we can also write out the following thermodynamic relations in L-set,

$$C_{V,\bar{N}^{(3)}} = \left(\frac{\partial U_q^{(3)}}{\partial T}\right)_{V,\bar{N}^{(3)}} = T\left(\frac{\partial S_q}{\partial T}\right)_{V,\bar{N}^{(3)}}, \tag{99}$$

$$\left(\frac{\partial U_q^{(3)}}{\partial V}\right)_{T,\ \bar{N}^{(3)}} = T\left(\frac{\partial P}{\partial T}\right)_{V,\bar{N}^{(3)}} - P, \tag{100}$$

$$\left(\frac{\partial U_q^{(3)}}{\partial \bar{N}^{(3)}}\right)_{T,\ V} = \mu - T\left(\frac{\partial \mu}{\partial T}\right)_{V,\bar{N}^{(3)}}, \tag{101}$$

$$C_{P,\bar{N}^{(3)}} = \left(\frac{\partial H_q^{(3)}}{\partial T}\right)_{P,\bar{N}^{(3)}} = T\left(\frac{\partial S_q}{\partial T}\right)_{P,\bar{N}^{(3)}}, \tag{102}$$

$$\left(\frac{\partial H_q^{(3)}}{\partial P}\right)_{T,\bar{N}^{(3)}} = V - T\left(\frac{\partial V}{\partial T}\right)_{P,\bar{N}^{(3)}}, \tag{103}$$

$$\left(\frac{\partial H_q^{(3)}}{\partial \bar{N}^{(3)}}\right)_{T,P} = \mu - T\left(\frac{\partial \mu}{\partial T}\right)_{P,\bar{N}^{(3)}}, \tag{104}$$

$$C_{P,\bar{N}^{(3)}} - C_{V,\bar{N}^{(3)}} = T\left(\frac{\partial P}{\partial T}\right)_{V,\bar{N}^{(3)}}\left(\frac{\partial V}{\partial T}\right)_{P,\bar{N}^{(3)}}. \tag{105}$$

In the next section, we will discuss the correlations of these nonextensive thermodynamic relations derived from different set of formalisms.

## 6. The correlations between the nonextensive thermodynamic relations for open systems

It is interesting to check the correlations between the nonextensive thermodynamic relations from different set of formalisms. From the fundamental thermodynamic equations about the P-internal energy and the L-internal energy, that is, the (79) and (80), we can find that

$$T_q = \left( \frac{\partial U_q^{(2)}}{\partial S_q} \right)_{V, \bar{N}^{(2)}} = c_q \left( \frac{\partial U_q^{(3)}}{\partial S_q} \right)_{V, \bar{N}^{(3)}}, \tag{106}$$

$$-P_q = \left( \frac{\partial U_q^{(2)}}{\partial V} \right)_{S_q, \bar{N}^{(2)}} = c_q \left( \frac{\partial U_q^{(3)}}{\partial V} \right)_{S_q, \bar{N}^{(3)}}, \tag{107}$$

$$\mu = \left( \frac{\partial U_q^{(2)}}{\partial \bar{N}^{(2)}} \right)_{S_q, V} = \left( \frac{\partial U_q^{(3)}}{\partial \bar{N}^{(3)}} \right)_{S_q, V}. \tag{108}$$

From (81) and (82), we get that

$$-P_q = \left( \frac{\partial F_q^{(2)}}{\partial V} \right)_{T_q, \bar{N}^{(2)}} = c_q \left( \frac{\partial F_q^{(3)}}{\partial V} \right)_{T, \bar{N}^{(3)}}, \tag{109}$$

$$-S_q = \left( \frac{\partial F_q^{(2)}}{\partial T_q} \right)_{V, \bar{N}^{(2)}} = \left( \frac{\partial F_q^{(3)}}{\partial T} \right)_{V, \bar{N}^{(3)}}, \tag{110}$$

$$\mu = \left( \frac{\partial F_q^{(2)}}{\partial \bar{N}^{(2)}} \right)_{T_q, V} = \left( \frac{\partial F_q^{(3)}}{\partial \bar{N}^{(3)}} \right)_{T, V}. \tag{111}$$

Moreover, from (83) and (84), we obtain that

$$T_q = \left( \frac{\partial H_q^{(2)}}{\partial S_q} \right)_{P_q, \bar{N}^{(2)}} = c_q \left( \frac{\partial H_q^{(3)}}{\partial S_q} \right)_{P, \bar{N}^{(3)}}, \tag{112}$$

$$V = \left( \frac{\partial H_q^{(2)}}{\partial P_q} \right)_{S_q, \bar{N}^{(2)}} = c_q \left( \frac{\partial H_q^{(3)}}{\partial P} \right)_{S_q, \bar{N}^{(3)}}, \tag{113}$$

$$\mu = \left( \frac{\partial H_q^{(2)}}{\partial \bar{N}^{(2)}} \right)_{S_q, P_q} = \left( \frac{\partial H_q^{(3)}}{\partial \bar{N}^{(3)}} \right)_{S_q, P}. \tag{114}$$

Then from (85) and (86), we gain that

$$-S_q = \left( \frac{\partial G_q^{(2)}}{\partial T_q} \right)_{P_q, \bar{N}^{(2)}} = \left( \frac{\partial G_q^{(3)}}{\partial T} \right)_{P, \bar{N}^{(3)}}, \tag{115}$$

$$V = \left( \frac{\partial G_q^{(2)}}{\partial P_q} \right)_{T_q, \bar{N}^{(2)}} = \left( \frac{\partial G_q^{(3)}}{\partial P} \right)_{T, \bar{N}^{(3)}}, \tag{116}$$

$$\mu = \left( \frac{\partial G_q^{(2)}}{\partial \bar{N}^{(2)}} \right)_{T_q, P_q} = \left( \frac{\partial G_q^{(3)}}{\partial \bar{N}^{(3)}} \right)_{T, P} . \tag{117}$$

Lastly, from (87) and (88), we can get that

$$-P_q = \left( \frac{\partial J_q^{(2)}}{\partial V} \right)_{T_q, \mu} = c_q \left( \frac{\partial J_q^{(3)}}{\partial V} \right)_{T, \mu} , \tag{118}$$

$$-S_q = \left( \frac{\partial J_q^{(2)}}{\partial T_q} \right)_{V, \mu} = \left( \frac{\partial J_q^{(3)}}{\partial T} \right)_{V, \mu} , \tag{119}$$

$$-\bar{N}^{(2)} = \left( \frac{\partial J_q^{(2)}}{\partial \mu} \right)_{V, T_q} = c_q \left( \frac{\partial J_q^{(3)}}{\partial \mu} \right)_{V, T} . \tag{120}$$

From the correlations presented above, an interesting equivalent rule is found that, in the partial derivative calculations with constant entropy or volume, the Tsallis factor is an invariable, and it can be directly extracted from the partial derivative. By use of this rule, more correlations can be found. It is not necessary to list all the correlations. We just give several examples. From (99), it can be seen that the heat capacity with constant volume satisfies

$$C_{V, \bar{N}^{(3)}} = \left( \frac{\partial U_q^{(3)}}{\partial T} \right)_{V, \bar{N}^{(3)}} = \left( \frac{\partial U_q^{(2)}}{\partial T_q} \right)_{V, \bar{N}^{(2)}} = C_{Vq, \bar{N}^{(2)}} \equiv C_V . \tag{121}$$

From (100), we have

$$\left( \frac{\partial U_q^{(2)}}{\partial V} \right)_{T_q, \ \bar{N}^{(2)}} = c_q \left( \frac{\partial U_q^{(3)}}{\partial V} \right)_{T, \ \bar{N}^{(3)}} , \tag{122}$$

and from (101) we get

$$\left( \frac{\partial U_q^{(2)}}{\partial \bar{N}^{(2)}} \right)_{T_q, \ V} = \left( \frac{\partial U_q^{(3)}}{\partial \bar{N}^{(3)}} \right)_{T, \ V} . \tag{123}$$

By the application of the rule, it is easy to find the statistical expressions of both the L-internal energy from the (10),

$$U_q^{(3)} = -\frac{\partial}{\partial \beta} \ln_q Z_q^{(2)} , \tag{124}$$

and the L-number from the (11),

$$\bar{N}^{(3)} = -\frac{\partial}{\partial \gamma} \ln_q Z_q^{(2)} . \tag{125}$$

Moreover, by recourse to the links of (50), (56), (71) and (77), the expressions of L-free energy and L-grand potential are given respectively by

$$F_q^{(3)} = -kT \ln_q Z_q^{(2)} + \mu \bar{N}^{(3)} , \tag{126}$$

$$J_q^{(3)} = -kT \ln_q Z_q^{(2)} . \tag{127}$$

Based on the above two expressions, from (109) or (118) the expression of L-pressure is easily obtained,

$$P = kT [Z_q^{(2)}]^{1-q} \frac{\partial \ln Z_q^{(2)}}{\partial V} . \tag{128}$$

It is difficult to calculate the grand partition function for the open systems in the above equation, yet according to the case in a closed system [20], we have the following approximation for a nonextensive gas,

$$Z_q^{(2)} = a(\bar{N}^{(3)}, q) \ V^{\bar{N}^{(3)}} (kT_q)^{\frac{3}{2}\bar{N}^{(3)}}, \qquad (129)$$

where $a$ is an unknown function of the L-number and the parameter $q$. Then we get

$$PV = \bar{N}^{(3)} kT[Z_q^{(2)}]^{1-q}, \qquad (130)$$

and further,

$$P_q V = \bar{N}^{(3)} kT_q [Z_q^{(2)}]^{1-q}. \qquad (131)$$

Similarly, for the nonextensive gas, the P-internal energy is

$$U_q^{(2)} = \frac{3}{2} \bar{N}^{(3)} kT_q [Z_q^{(2)}]^{1-q}, \qquad (132)$$

and the L-internal energy is

$$U_q^{(3)} = \frac{3}{2} \bar{N}^{(3)} kT [Z_q^{(2)}]^{1-q}. \qquad (133)$$

It can be seen that for the heat capacities there are the relations,

$$C_V \equiv C_{Vq,\bar{N}^{(2)}} = C_{V,\bar{N}^{(3)}} = \frac{3}{2} \bar{N}^{(3)} k [Z_q^{(2)}]^{1-q} [1 + \frac{3}{2} \bar{N}^{(3)} (1-q)]. \qquad (134)$$

Considering the heat cut-off of the power-law $q$-distribution functions (9) and (14), there is $1-q>0$; then the heat capacity in the above equation is always positive for a normal nonextensive gas. The further work would be carried out in other papers. Here we do not give more discussions.

## 7. Conclusions and discussions

In this paper, we further studied the nonextensive thermodynamics for the open systems based on the maximum entropy principle. The temperature duality assumption is adopted and all the nonextensive thermodynamic relations and thermodynamic quantities are endowed with the dual interpretations. By doing this, a nonextensive thermodynamic formalism consisting of two sets of parallel Legendre transformations is proposed.

For an open system, the particle number is a state variable, which should be defined in the two sets of the nonextensive formalisms. So we presented two particle number definitions: one is called P-number and the other is called L-number, which can be linked through the Tsallis factor. The physical meanings of these two particle numbers are clearly given in our treatments. For an open complex system, the L-number is the number of the non-creation particles restricted to the system, while the P-number is the sum of the numbers of three kinds of particles: the non-creation particles and creation particles inside the system, and the particles also existing on the system surface, which take part in the interactions between the system and its surroundings. Of course, the creation particles can also be annihilated due some unknown reasons. Anyway, according to (12), because the partition function in P-set of formalism is more than unity and also $1-q>0$, there should be

$$c_q > 1, \quad \bar{N}^{(2)} > \bar{N}^{(3)}. \qquad (135)$$

It is reasonable for the nonadditivity of P-number, now that it includes the number of creation particles.

In our treatments, the two sets of nonextensive formalisms share the common entropy, volume and chemical potential. Apart from this, the chemical potential also participates into the chemical balance between different subsystems or phases. This makes the chemical potential more special that it is the only one intensive quantity that does not require the dual interpretations. This is rational because chemical reactions all occur in local regions of the open complex systems.

All the Legendre transformations in this paper are directly originated from the traditional thermodynamics. This means the thermodynamic relations in the nonextensive thermodynamics are identical in forms to those in traditional thermodynamics. The main difference is that in the proposed treatments there exist two sets of parallel Legendre transformation structures, which are linked through the Tsallis factor.

The Maxwellian relations and other thermodynamic relations are easily deduced in the proposed nonextensive thermodynamic formalism consisting of particle numbers as state variables. By studying the basic nonextensive thermodynamic equations, these correlations between the nonextensive thermodynamic relations within different set of formalisms are found. And an equivalent rule is deduced that in the calculations of partial derivative with constant volume or entropy, the Tsallis factor is invariable and it can be extracted from the partial derivative.

By use of the equivalent rule, more correlations are found and these two sets of nonextensive formalisms share the same heat capacity (see (121)). These expressions of internal energies are easily obtained for a nonextensive gas. Also with the rule, the heat capacity of the nonextensive gas is always positive. However, the equivalent rule might be not valid in some situations, such as the nonexensive systems with small size, or the self-gravitating systems, where the long range interactions are regarded as some kind of "phase transition", in which the heat capacity might be negative.

## Acknowledgements

Zheng Y. wishes to express thanks to the colleges at the Henan Institute of Technology and the Xinxiang University in Xinxiang city. This work is supported by the National Natural Science Foundation of China under Grant No. 11775156, and it is partially supported by the National Natural Science Foundation of China under Grant No. 11405092.

## Appendix

In ordr to verify (15), we shoulde prove that

$$\bar{Z}_q^{(3)} = \sum [1 - \frac{(1-q)}{c_q} [\beta(\varepsilon_i' - U_q^{(3)}) + \gamma(N_i - \bar{N}^{(3)})]]^{q/(1-q)} . \tag{A.1}$$

From the equation,

$$\beta U_q^{(3)} + \gamma \bar{N}^{(3)} = \frac{\sum p_i^q (\beta \varepsilon_i' + \gamma N_i)}{c_q} , \tag{A.2}$$

we have that

$$0 = \frac{\sum p_i^q (\beta \varepsilon_i' + \gamma N_i - \beta U_q^{(3)} - \gamma \bar{N}^{(3)})}{c_q} = \frac{1}{1-q} \sum [1 - \frac{(1-q)}{c_q} [\beta(\varepsilon_i' - U_q^{(3)}) + \gamma(N_i - \bar{N}^{(3)})]]^{q/(1-q)}$$

$$- \frac{1}{1-q} \sum [1 - \frac{(1-q)}{c_q} [\beta(\varepsilon_i' - U_q^{(3)}) + \gamma(N_i - \bar{N}^{(3)})]]^{1/(1-q)} . \tag{A.3}$$

Because that

$$\bar{Z}_q^{(3)} \equiv \sum [1 - \frac{(1-q)}{c_q} [\beta(\varepsilon_i' - U_q^{(3)}) + \gamma(N_i - \bar{N}^{(3)})]]^{1/(1-q)} , \tag{A.4}$$

the (A.1) and (15) are verified.

## References

[1] Tsallis, C. J. Stat. Phys. **52**, 479 (1988).

[2] Du, J. L. Europhys. Lett. **67**, 893 (2004).

[3] Leubner, M.P. Astrophys. J. **632**, L1 (2005).

[4] Du, J. L. New Astron. **12**, 60 (2006).

[5] Du, J. L. Astrophys. Space Sci. **312**, 47 (2007) and the references therein.

[6] Zheng, Y. and Du, J. : Two physical explanations of the nonextensive parameter in a self-gravitating system. EPL **107**, 60001 (2014)

[7] Lima, J.A.S., Silva Jr., R., Santos, J. Phys. Rev. E **61**, 3260 (2000).

[8] Du, J. L. Phys. Lett. A **329**, 262 (2004).

[9] Liu, L.Y., Du, J. L. Physica A **387**, 4821 (2008).

[10] Yu, H. N., Du, J. L. Ann. Phys. **350**,302 (2014).

[11] Du, J. L. Physica A **391**, 1718 (2012).

[12] Yin, C. T., Du, J. L. Physica A **395**, 416 (2014).

[13] Oikonomou, T., Provata, A., Tirnakli, U. Physica A **387**, 2653 (2008).

[14] Rolinski, O. J., Martin, A., Birch, D. J. S. Ann. N. Y. Acad. Sci. **1130**, 314 (2008).

[15] Eftaxias, K., Minadakis, G., Potirakis, S.M., Balasis, G. Physica A **392**, 497 (2013).

[16] Abe, S., Martınez, S., Pennini, F., Plastino, A. Phys. Lett. A **281**, 126 (2001).

[17] Toral R. Physica A **317**, 209 (2003).

[18] Abe S. Physica A **368**, 430 (2006).

[19] Guo, L. N., Du, J. L. Physica A **390**, 183 (2011).

[20] Zheng Y., Du J. Continuum Mech. Thermodyn. **28**, 1791 (2016).

[21] E. M. F. Curado and C. Tsallis, J. Phys. A **24**, L69 (1991).

[22] C. Tsallis, R. S. Mendes, and A. R. Plastino, Physica A **261**,534 (1998).

[23] A. R. Plastino and A. Plastino, Phys. Lett. A **177**, 177(1993).

[24] Wang Q A, Le Méhauté A. Chaos, Solitons & Fractals **15**, 537 (2003).

[25] Zheng Y, Du J, Liang F. Continuum Mech. Thermodyn. **30**, 629 (2018).